\documentclass[12pt, preprint]{aastex}
\usepackage{graphicx}

\def\ftd{F_{\rm TD}}
\def\ratio{(R_{\rm app}/D)^2}

\begin{document}

\title{The Mass and Radius of the Neutron Star in EXO~1745$-$248}

\author{Feryal \"Ozel\altaffilmark{1}, Tolga G\"uver and Dimitrios
Psaltis\altaffilmark{1}}

\affil{University of Arizona, Department of Astronomy, 933 N. Cherry 
Ave., Tucson, AZ 85721}

\altaffiltext{1}{Steward Observatory, University of Arizona} 

\begin{abstract}

Bursting X-ray binaries in globular clusters are ideal sources for
measuring neutron star masses and radii, and hence, for determining
the equation of state of cold, ultradense matter.  We use
time-resolved spectroscopic data from EXO~1745$-$248 during
thermonuclear bursts that show strong evidence for photospheric radius
expansion to measure the Eddington flux and the apparent surface area
of the neutron star. We combine this with the recent measurement of
the distance to the globular cluster Terzan~5, where this source
resides, to measure the neutron star mass and radius. We find tightly
constrained pairs of values for the mass and radius, which are
centered around $M=1.4~M_\odot$ and $R=11$~km or around $M=1.7~M_\odot$ 
and $R=9$~km. These values favor nucleonic equations of state
with symmetry energy that is relatively low and has a weak dependence
on density.

\end{abstract}


\section{Introduction}

Accreting neutron stars that show thermonuclear X-ray bursts are
optimal sources for determining the equation of state of cold,
ultradense matter. They exhibit a number of spectroscopic phenomena
that depend on the neutron star mass and radius, which can be used to
measure these fundamental stellar properties. In particular, the
apparent surface area of thermal emission during the cooling tail of a
thermonuclear burst, as well as the peak flux achieved during strong,
so-called photospheric radius expansion bursts, which reach the
Eddington limit, provide two such observable quantities (van Paradijs
1978, 1979; see also Lewin, van Paradijs, \& Taam 1993).

The large collecting area and the systematic monitoring of a number of
X-ray bursters by the Rossi X-ray Timing Explorer (RXTE) has generated
a large database of high quality observations that we use in this
study.  The peak flux achieved in numerous photospheric radius
expansion bursts from several sources has been shown to remain nearly
constant, within a range as narrow as a few percent in 4U~1728$-$34
(Galloway et al.\ 2003), and the apparent surface areas to be
reproducible between bursts (Galloway et al.\ 2008). This provides
observational justification to the theoretical expectation that, in
low magnetic field neutron stars, thermonuclear bursts not only
quickly engulf the entire stellar surface but also that strong bursts
reach an intrinsic limit associated with the Eddington luminosity.

These two spectroscopic phenomena can be combined either with another
spectroscopic measurement, such as a surface redshift (Cottam,
Paerels, \& Mendez 2002; see also Cottam et al.\ 2008) or with an
accurate, independent distance to the neutron star, to break the
intrinsic degeneracies in the neutron star properties and determine
both the mass and the radius of the neutron star, independently
(\"Ozel 2006).  X-ray bursters in globular clusters are unique in this
respect, since the distances to the clusters can be independently
measured. In this paper, we use the thermonuclear burst data of the
source EXO~1745$-$248 located in Terzan~5 to measure the mass and
radius of the neutron star.

The low mass X-ray binary EXO~1745$-$248 was discovered with Hakucho
in August 1980 (Makishima et al.\ 1981). The source showed Type-I
X-ray bursts, with intervals as short as 8 min (Inoue et al.\ 1984).
It was again detected in 2000 during a RXTE/PCA scan of the galactic
bulge as a transient X-ray burster (Markwardt \& Swank 2000).
EXO~1745$-$248 was also observed by the Chandra X-ray Observatory in
2000 and 2003. Heinke et al.\ (2003) used the Chandra and RXTE
observations to suggest that the source is an ultracompact binary and
also identified a possible infrared counterpart in the Hubble Space
Telescope (HST) images of the cluster. No burst oscillations have been
reported from this source (Galloway et al.\ 2008).

Terzan 5 is one of the most metal rich globular cluster in the galaxy,
with a metallicity close to solar (Origlia et al.\ 2004).  Recently,
Ortolani et al.\ (2007) revisited the distance measurements to
Terzan~5 using HST/NICMOS data. Using NICMOS instrumental magnitudes
and two separate reddening laws (Schlegel et al. 1998; Lee et al.\
2001) to obtain the infrared extinction slope in the instrumental
bands, they measured a distance of 6.3~kpc to this cluster. The two
main sources of error in this measurement are related to uncertainties
in the color and magnitude measurements of the HB of the cluster, as
well as to metallicity uncertainties, while the slope of the reddening
law introduces a much smaller error. The combined error of 0.2 mag
corresponds to a distance uncertainty of about 10\%, which we will
adopt here. Note that Ortolani et al.\ (2007) also used two
calibrations for the conversion of the NICMOS to JHK magnitudes, which
resulted in different values for the distance. Due to the significant
width and the large displacement of the NICMOS F110W filter compared
to the ground-based J filter, the transformations between these bands
are color dependent and hence, suffer from systematic uncertainties
that are difficult to quantify. For this reason, we will only use the
distance measurement obtained from NICMOS instrumental magnitudes.

In this paper, we combine the distance measurement to Terzan~5 with
the observations of radius expansion bursts obtained by RXTE to
determine the mass and the radius of the neutron star in
EXO~1745$-$248. In Section 2, we analyze the Eddington limited bursts
from this source. In Section 3, we use these observations to determine
the mass and radius of the neutron star and describe the formal method
for assessing the uncertainties in the measurements using this
technique. In Section 4, we discuss our results and compare them to
several equations of state for neutron star matter.

\section{Spectral Analysis of X-ray Bursts}

EXO~1745$-$248 has been observed with RXTE for 148~ks, during which
two Type-I X-ray bursts were discovered, with clear evidence for
photospheric radius expansion (Galloway et al.\ 2008).  Note that,
while Galloway et al.\ (2008) identified 20 more candidate non-PRE
bursts that satisfied their trigger criteria, spectral analyses of
those bursts revealed that the distinctive cooling associated with
Type~I bursts did not occur in these events, strongly suggesting that
they are Type~II instead (Lewin et al.\ 1993). 

In order to analyze the PRE bursts, we extracted time resolved
2.5$-$25.0 keV X-ray spectra using the ftool {\em seextrct} and
included the data from all the RXTE/PCA layers. We used the Science
Event mode data with the E\_125$\mu$s\_64M\_0\_1s configuration, which has
a nominal time resolution of 125~$\mu$s in 64 spectral channels.  We
binned the X-ray spectra in 27 spectral channels and over 0.25~s (for
count rates above 6000 ct~s$^{-1}$) and over 0.5~s (for count rates
between 3000 and 6000 ct~s$^{-1}$) time intervals during each
burst. Following Galloway et al.\ (2008), we extracted a 16~s spectrum
prior to the burst and used it as background. We generated separate
response matrix files for each burst using the PCARSP version 10.1 and
took into account the offset pointing of the PCA during the creation
of the response matrix files.

We fit the extracted spectra with a blackbody function, using the
hydrogen column density value of N$_{\rm H} = 1.4 \times
10^{22}$~cm$^{-2}$ determined by Wijnands et al.\ (2005) from Chandra
observations. We used XSPEC v12 (Arnaud 1996) for our spectral
analysis. For each spectrum, we calculated bolometric fluxes using
equation~(3) of Galloway et al.\ (2008). Figure~1 shows an example
countrate spectrum as well as the best fit blackbody model. There are
no systematic residuals in the fit, and the addition of any other
spetcral components (e.g., a power-law model) is not statistically
significant. 

In Figure~2, we show the distribution of the $\chi^2$/d.o.f. values
that we obtained by fitting the X-ray spectra of the source during the
2 PRE bursts and compare it to the expected distribution for 25
degrees of freedom. All fits with $\chi^2/{\rm d.o.f.} < 1.5$ follow
the expected distribution and are, therefore, statistically
acceptable.  However, the five spectra with $\chi^2/{\rm d.o.f.} >
1.5$ are outliers, which are likely to be dominated by systematic
uncertainties.  We rejected these fits from the subsequent analyses.

We show in Figure~3 the bolometric flux, the blackbody temperature,
and the normalization of the model spectra during the evolution of the
two PRE bursts. The characteristic decrease of the temperature and the
increase of the photospheric radius around the burst peak, as well as
the cooling of the burst emission at a constant photospheric radius
for both bursts can be seen in both bursts.

In PRE bursts, the Eddington limit at the surface of the neutron star
is thought to correspond to the point in each burst when the
normalization of the blackbody gets its lowest value while the
temperature reaches its highest (Damen et al.\ 1990). The spectral
properties of the two PRE bursts during this so-called touchdown point
are consistent with each other, as demonstrated in
Figure~\ref{touchdown}. The combined best-fit value for the touchdown
flux between the two bursts is $(6.25 \pm 0.20) \times
10^{-8}$~erg~cm$^{-2}$~s$^{-1}$. Note that the ratios of the peak to
touchdown fluxes in these two bursts are well within the value
expected from the general relativistic effects alone, and therefore,
this source is not subject to the bias discussed in Galloway, \"Ozel,
\& Psaltis (2008).

The second observational quantity that we determine from the spectral
fits is the apparent radius of the emitting region during the cooling
phase of the bursts. This is given directly by the normalization of
the blackbody function, $A \equiv (R_{\rm app}/D)^2$, where $R_{\rm
app}$ is the radius corresponding to the apparent emitting surface
area and $D$ is the distance to the source. We chose the intervals
4.5$-$15~s in both bursts, during which the apparent radius is
constant. Fitting the cooling tails of these bursts individually
resulted in values for the ratio $A = 104.0 \pm 1.0$~km$^2$~kpc$^{-2}$
and $A = 130.0 \pm 1.0$~km$^2$~kpc$^{-2}$. Similar systematic
uncertainties have been observed in the Eddington fluxes from PRE
bursts from, e.g., 4U~1728$-$34 (Galloway et al.\ 2003) and have been
attributed to the variable reflection off of the accretion disk that
changes at a superorbital period. Such a phenomenon can introduce
systematic uncertainties in the apparent surface areas measured during
the cooling tails of bursts. Because the systematic errors dominate
over the statistical errors in this particular case, we will assume a
boxcar probability distribution over this quantity in the range $A =
116 \pm 13$~km$^2$~kpc$^{-2}$.

\section{Determination of the Neutron Star Mass and Radius}

In an approach similar to \"Ozel (2006), we use the spectroscopic
measurements of the touchdown flux $\ftd$ and the ratio $A$ during the
cooling tails of the bursts, together with the measurement of the
distance $D$ to the source in order to determine the neutron star mass
$M$ and radius $R$. The observed spectroscopic quantities depend on
the stellar parameters according to the relations
\begin{equation}
\ftd=\frac{GMc}{k_{\rm es}D^2}\left(1-\frac{2GM}{Rc^2}\right)^{1/2}
\end{equation}
and
\begin{equation}
A=\frac{R^2}{D^2f_{\rm c}^4}\left(1-\frac{2GM}{Rc^2}\right)^{-1}\;,
\end{equation}
where $G$ is the gravitational constant, $c$ is the speed of light,
$k_{\rm es}$ is the opacity to electron scattering, and $f_{\rm c}$ is
the color correction factor. 

In the absence of errors in the determination of the observable
quantities, the last two equations can be solved for the mass and
radius of the neutron star. However, because of the particular
dependences of $\ftd$ and $A$ on the neutron star mass and radius (see
also Fig.~1 in \"Ozel 2006), the loci of mass-radius points that
correspond to each observable intersect, in general, at two distinct
positions. Moreover, the diverse nature of uncertainties associated to
each of the observables requires a formal assessment of the
propagation of errors, which we present here.

We assign a probability distribution function to each of the
observable quantities and denote them by $P(D)dD$, $P(\ftd)d\ftd$, and
$P(A)dA$. Because the various measurements that lead to the
determination of the three observables are independent of each other,
the total probability density is simply given by the product
\begin{eqnarray} 
&&P(D,\ftd,A)dDd\ftd dA =\nonumber\\ &&\qquad \qquad
P(D)P(\ftd)P(A) dD d\ftd dA\;.  
\label{eq:firstdistrib} 
\end{eqnarray}

Our goal is to convert this probability density into one over the
neutron-star mass, $M$, and radius, $R$. We will achieve this by
making a change of variables from the pair $(\ftd,A)$ to $(M,R)$ and
then by marginalizing over distance. Formally, this implies that
\begin{eqnarray}
&& P(D,M,R)dDdMdR = 
\frac{1}{2} P(D)P[\ftd(M,R,D)]
\nonumber \\
&& \qquad P[A(M,R,D)] J\left(\frac{\ftd,A}{M,R}\right) dD dMdR\;,
\label{eq:transform}
\end{eqnarray}
where $J(\ftd,A/M,R)$ is the Jacobian of the transformation. It is
important to emphasize here that, given a distance D, not all pairs of
the observables $(\ftd,A)$ can be obtained with real values for the
neutron-star mass and radius. For this reason, the final distribution
will not be normalized, even if the three distributions of
equation~(\ref{eq:firstdistrib}) are. In addition, the factor $1/2$
appears in equation~(\ref{eq:transform}) because nearly all pairs of
the observables $(\ftd,A)$ correspond to two distinct pairs of
$(M,D)$.  There is only a region of the parameter space for which the
pair of observables corresponds to a single pair of values for the
mass and radius. However, this region has zero volume and, therefore,
will not contribute to the final probability distribution after we
marginalize over distance.

We can now use the above expressions to calculate the Jacobian of the 
transformation
\begin{eqnarray}
J\left(\frac{\ftd,A}{M,R}\right)&=&\frac{GcR}{f_{\rm c}^4 k_{\rm es}D^4}
\left[1-\frac{6GM}{Rc^2}+7\left(\frac{GM}{Rc^2}\right)\right] \nonumber \\
&& \left(1-\frac{2GM}{Rc^2}\right)^{-5/2}\;.
\label{eq:Jac}
\end{eqnarray}

For the source EXO~1745$-$248, the distance measurement is dominated by
systematic errors as discussed in \S1. We will, therefore, use a box-car 
probability distribution over distance, with a mean of $D_0=6.3$~kpc and 
a range of $\Delta D=0.1 D_0$, i.e., 
\begin{equation}
P(D)dD=\left\{\begin{array}{ll}
\frac{1}{\Delta D} & {\rm if}\quad\vert D-D_0\vert \le \Delta D/2\\
& \\
0 & {\rm otherwise}
\end{array}
\right.
\label{eq:PD}
\end{equation}

The measurements of the touchdown flux is consistent between the two
radius-expansion bursts (see Fig.~\ref{touchdown}) and is, therefore,
dominated only by statistical uncertainties. We assign a Gaussian
probability distribution for this quantity with a mean and a standard
deviation that we estimate by fitting a Gaussian function to the
product of the probability distributions that correspond to the
confidence contours shown in Figure~\ref{touchdown}. The result is a
mean of $F_{\rm 0}=6.25\times 10^{-8}$~erg~cm$^{-2}$~s$^{-1}$ and a
standard deviation of $\sigma_{\rm F}=0.2\times
10^{-8}$~erg~cm$^{-2}$~s$^{-1}$, i.e.,
\begin{equation}
P(\ftd)d\ftd=\frac{1}{\sqrt{2\pi \sigma_{\rm F}^2}}
\exp\left[-\frac{(\ftd-F_0)^2}{2\sigma_{\rm F}^2}\right]\;.
\label{eq:PF}
\end{equation}

Finally, the measurement of the ratio $A\equiv\ratio$ between the two
bursts is dominated by systematic uncertainties. We, therefore, assign
to this ratio a box car probability distribution with a mean of
$A_0=116$ and a range of $\Delta A=26$, i.e.,
\begin{equation}
P(A)dA=\left\{\begin{array}{ll}
\frac{1}{\Delta A} & {\rm if}\quad\vert A-A_0\vert \le \Delta A/2\\
& \\
0 & {\rm otherwise}
\end{array}
\right.
\label{eq:PA}
\end{equation}

The color correction factor that enters these expressions is
determined by models of burning neutron star atmospheres (e.g., Madej
et al.\ 2004). At the observed high temperatures of the bursts, as
well as in the absence of significant magnetic fields or heavy
elements (as evidenced by the lack of atomic transition lines in the
high resolution spectra), the Comptonized radiative equilibrium
atmosphere models can be reliably calculated. As discussed in \"Ozel
(2006), when the emerging flux is substantially sub-Eddington, as in
the case of the cooling tails of the bursts, the color correction
factor $f_{\infty}$ asymptotes to a value of $\simeq 1.4$, which we
adopt here. Finally, we use the electron scattering opacity
$\kappa_{es} = 0.20 (1+X)$~cm$^{2}$~g$^{-1}$, that depends on the
hydrogen mass fraction $X$.

We obtain the final distribution over neutron-star mass and radius by
inserting equations~(\ref{eq:Jac})--(\ref{eq:PA}) into
equation~(\ref{eq:transform}) and integrating over distance.
Figure~\ref{mr} shows the 1- and $2-\sigma$ contours for the mass and
radius of the neutron star in EXO~1745$-$248, for a hydrogen mass
fraction $X=0$. For larger values of the hydrogen mass fraction ($X
\gtrsim 0.1$), the masses and radii inferred individually from the
Eddington limit and the apparent surface area become rapidly
inconsistent with each other. This result is in line with the
identification of EXO~1745$-$248 with an ultracompact binary by Heinke
et al.\ (2003). Note that there are two distinct regions in the
mass-radius plane that are consistent with the data because of the
particular dependence of $\ftd$ and $A$ on the stellar mass and
radius, as discussed above.

\section{Discussion}

We used time-resolved spectroscopic data from EXO~1745$-$248 during
thermonuclear bursts that show strong evidence for photospheric radius
expansion to measure the Eddington flux and the apparent surface area
of the neutron star. We combined this with the recent measurement of
the distance to the globular cluster Terzan~5 (Ortolani et al.\ 2007),
where this source resides, to measure the neutron star mass and
radius. We found tightly constrained pairs of values for the mass and
radius, which are centered around $M=1.4~M_\odot$ and $R=11$~km or
around $M=1.7~M_\odot$ and $R=9$~km.

The confidence contours on the mass-radius plane (see Fig. 5) are in
best agreement with nucleonic equations of state without the presence
of condensates or strange matter. The leftmost family of mass-radius
relations is based on the assumption that the absolute ground state of
matter is made up of an approximately equal mixture of up, down, and
strange quarks. The primary difference between the other two families
of mass-radius relations is the symmetry properties of the equation of
state of neutron star matter. Moreover, the mass-radius relations with
deflection points are characteristic of calculations that incorporate
bosons that can condense and, thus, soften the equation of state at
high densities. The radius measurements presented here favor
relatively low values for the bulk symmetry energy with a weak density
dependence (see Lattimer \& Prakash 2001).

The measurement of the mass and radius of a neutron star can
significantly constrain the range of possibilities for the equation of
state of ultradense matter, as discussed above. However, it cannot
uniquely pinpoint to a single equation of state because of both the
measurement errors and the uncertainties in the fundamental parameters
that enter the nuclear physics calculations, such as the symmetry
energy of nucleonic matter or the bag constant for strange stars.
Further, even tighter constraints on the equation of state can be
obtained by combining observations of neutron stars with different
masses that will distinguish between the slopes of the predicted
mass-radius relations, which are determined entirely by the physics of
the neutron star interior.

A number of other constraints on neutron star radii have been obtained
to date using various methods. \"Ozel (2006) used spectroscopic
measurements of the Eddington limit and apparent surface area during
thermonuclear bursts, in conjuction with the detection of a redshifted
atomic line from the source EXO~0748$-$676, to determine a mass of $M
\ge 2.10 \pm 0.28~M_\odot$ and a radius $R \ge 13.8 \pm 1.8~{\rm km}$.
This radius measurement is consistent with the one presented in the
current paper to within 2$-\sigma$, and, therefore, several nucleonic
equations of state are consistent with both measurements. 

Radii have also been measured from globular cluster neutron stars in
binaries emitting thermally during quiescence, such as X7 in 47~Tuc
and others in $\omega$\ Cen, M~13, and NGC~2808 (Heinke et al.\ 2006;
Webb \& Barret 2007). (Note that we do not consider here isolated
neutron stars such as RX~J1856$-$3754 because of the unquantified
systematic uncertainties arising from the apparent temperature
anisotropies on the neutron star surfaces and their probable magnetic
nature; see Walter \& Lattimer 2002; Braje \& Romani 2002; Tiengo \&
Mereghetti 2007). These measurements have carved out large allowed
bands in the mass-radius plane, all of which are also consistent with
equations of state that predict neutron stars with radii $R \sim
11$~km. Future tight constraints on the masses and radii of additional
neutron stars with these and other methods (see, e.g., Lattimer \&
Prakash 2007) will resolve this long-standing question of high energy
astrophysics.

\acknowledgements

We thank Rodger Thompson for his help with understanding the NICMOS
calibrations, Duncan Galloway for his help with burst analyses,
Adrienne Juett for bringing the source to our attention, Jim Lattimer 
for sharing with us the mass-radius relations, and Martin
Elvis for useful conversations on constraining the neutron star
equation of state. We also thank an anonymous referee for useful
suggestions. F.\ \"O. acknowledges support from NSF grant
AST~07-08640. D.\ P. is supported by the NSF CAREER award NSF~0746549.

\begin{figure}
\centering
   \includegraphics[scale=0.5, angle=-90]{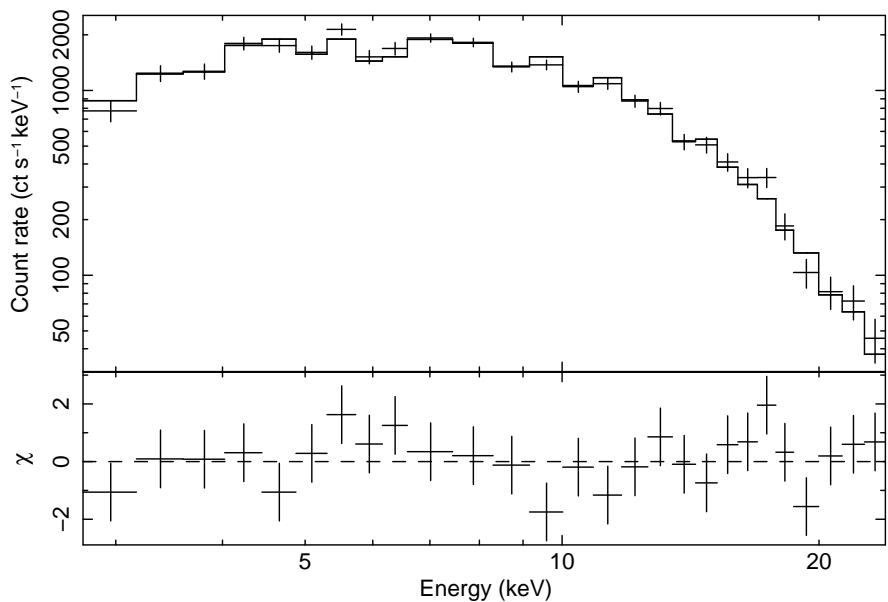}
\caption{An example count rate spectrum of EXO~1745$-$248 together 
with the best-fit blackbody model. The lower panel shows the residuals
of the fit, defined as $\chi = (x_i - x_m)/\sigma_i$, where $x_i$ and
$\sigma_i$ are the observed counts and uncertainty, respectively, in
the i-th spectral bin, and $x_m$ is the model prediction. This example 
corresponds to the touchdown point of the burst shown in the left 
panel of Figure~3.
}
\label{spectrum}
\end{figure}

\begin{figure}
\centering
   \includegraphics[scale=0.75]{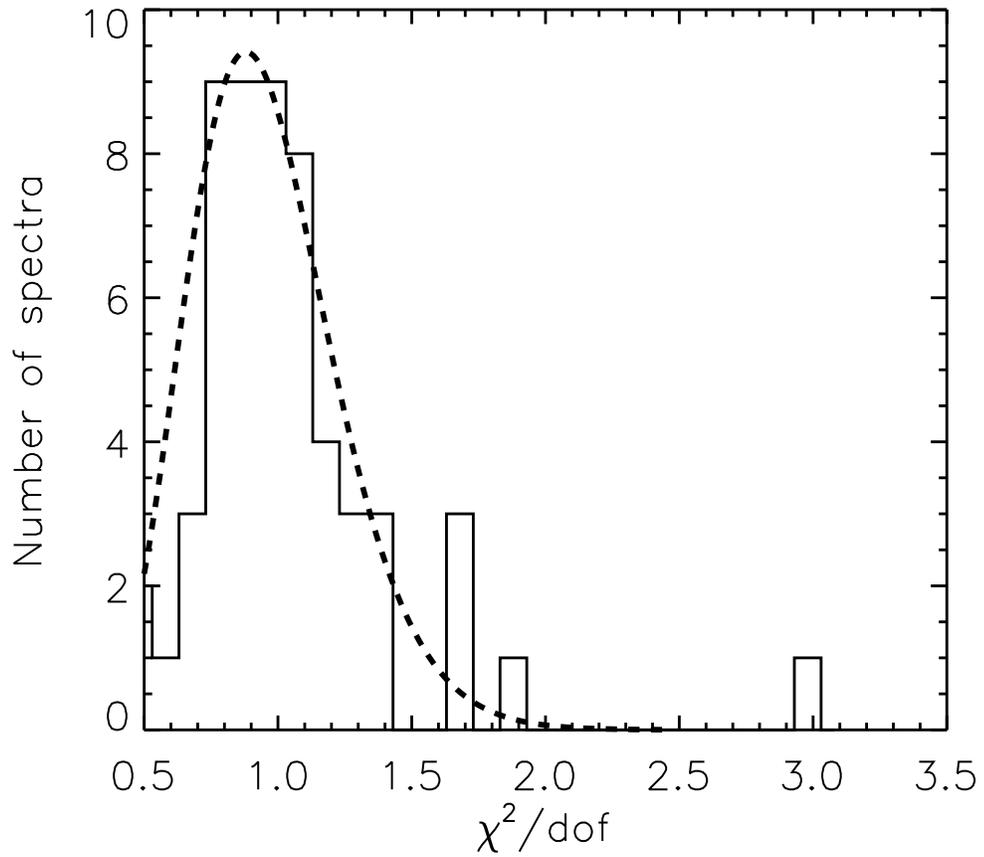}
\caption{The histogram shows the distribution of $\chi^2/{\rm d.o.f.}$ 
values obtained by fitting the spectra during the evolution of the two
photospheric radius bursts of EXO~1745$-$248. the dashed line shows
the expected $\chi^2/{\rm d.o.f.}$ distribution for 25 degrees of
freedom. The five spectral fits with $\chi^2/{\rm d.o.f.} > 1.5$ are
outliers and are excluded from the subsequent analyses. }
\label{chisq}
\end{figure}

\begin{figure*}
\centering
   \includegraphics[scale=0.4]{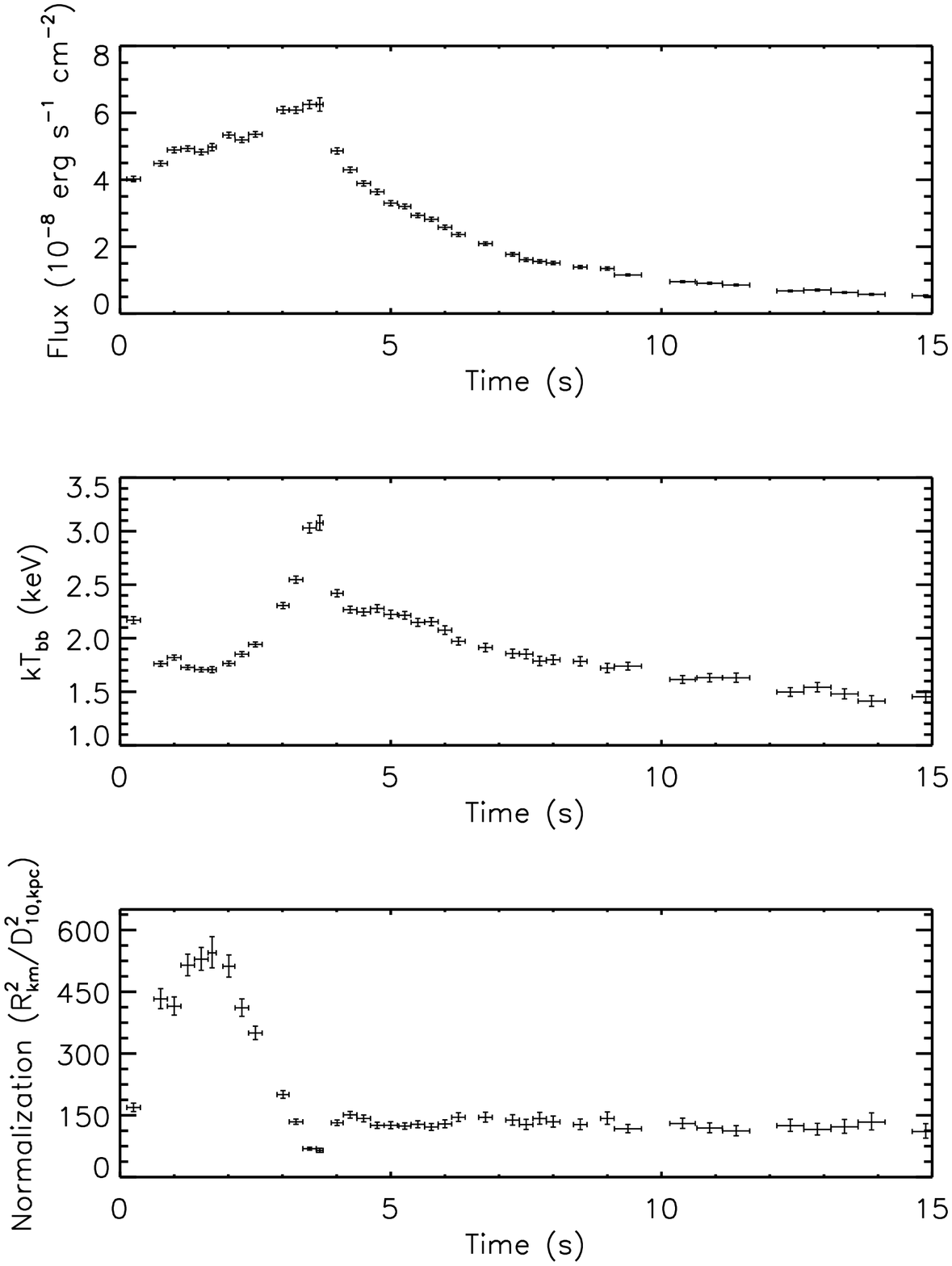}
   \includegraphics[scale=0.4]{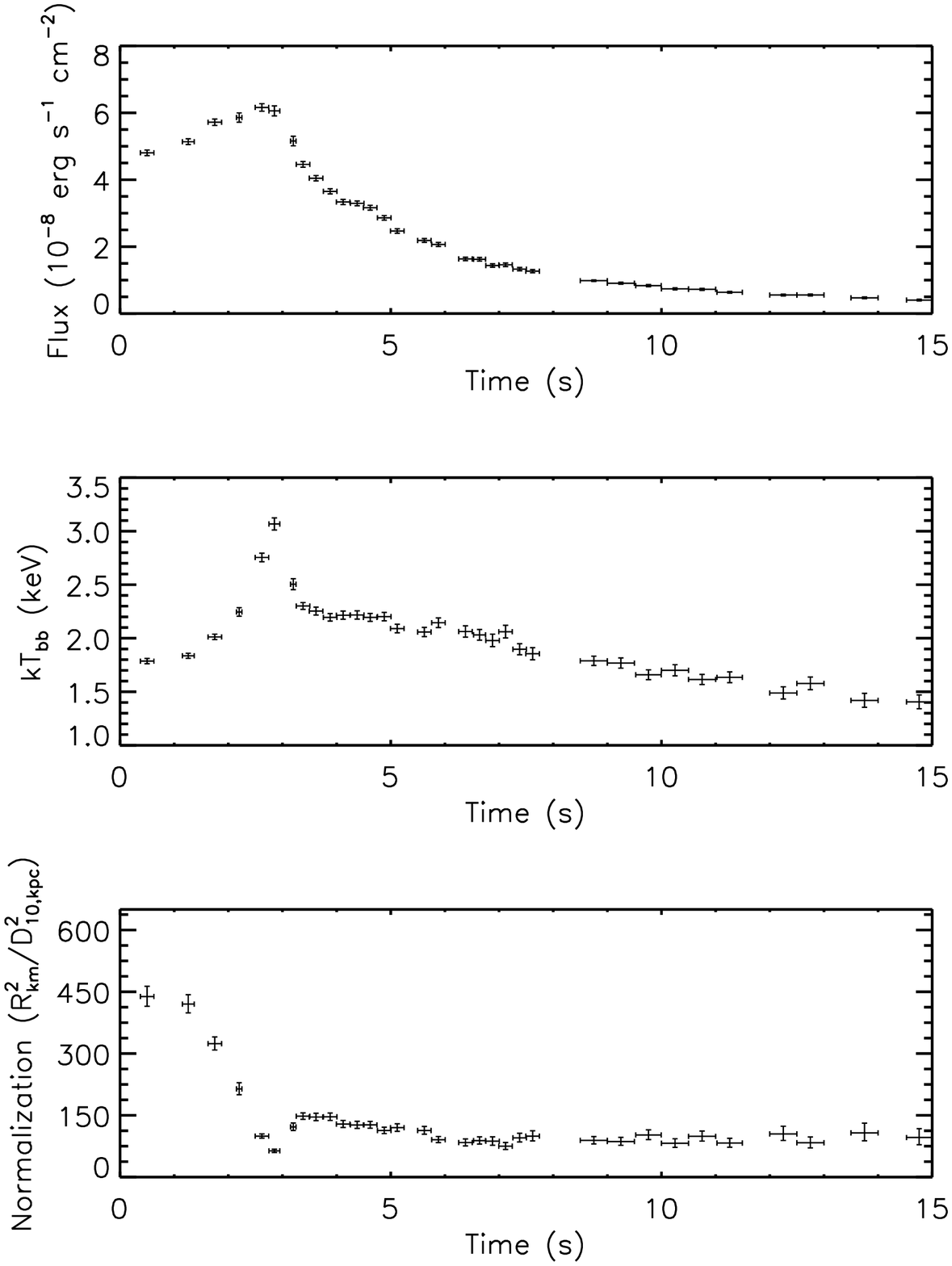}
\caption{The spectral evolution during the first 15 seconds of the two
Eddington limited thermonuclear bursts observed from EXO~1745$-$248 by
RXTE. The panels show the evolution of the flux, the blackbody
temperature and the apparent radius as observed at infinity, together
with their $1-\sigma$ statistical errors.}  \label{evolution}
\end{figure*}

\begin{figure}
\centering
   \includegraphics[scale=0.75]{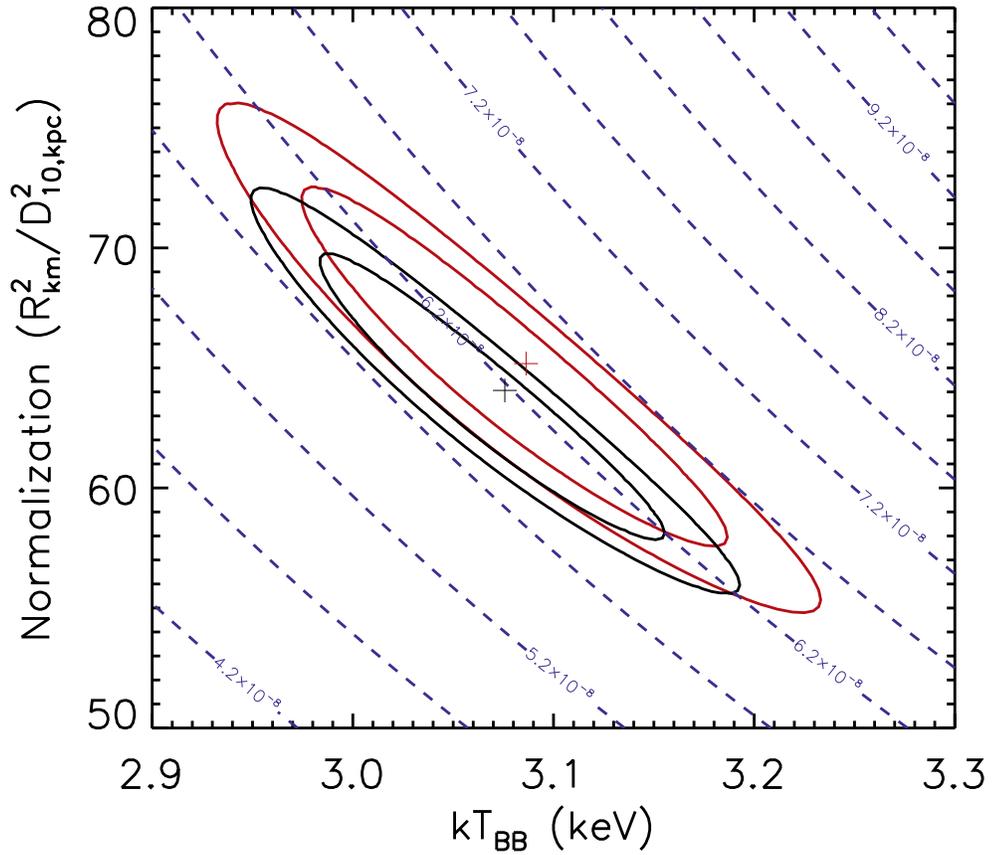}
\caption{The 1- and 2$-\sigma$ confidence contours of the
normalization and blackbody temperature obtained from fitting the two
PRE bursts during touchdown. The dashed lines show contours of
constant bolometric flux.}
\label{touchdown}
\end{figure}

\begin{figure}
\centering
   \includegraphics[scale=0.75]{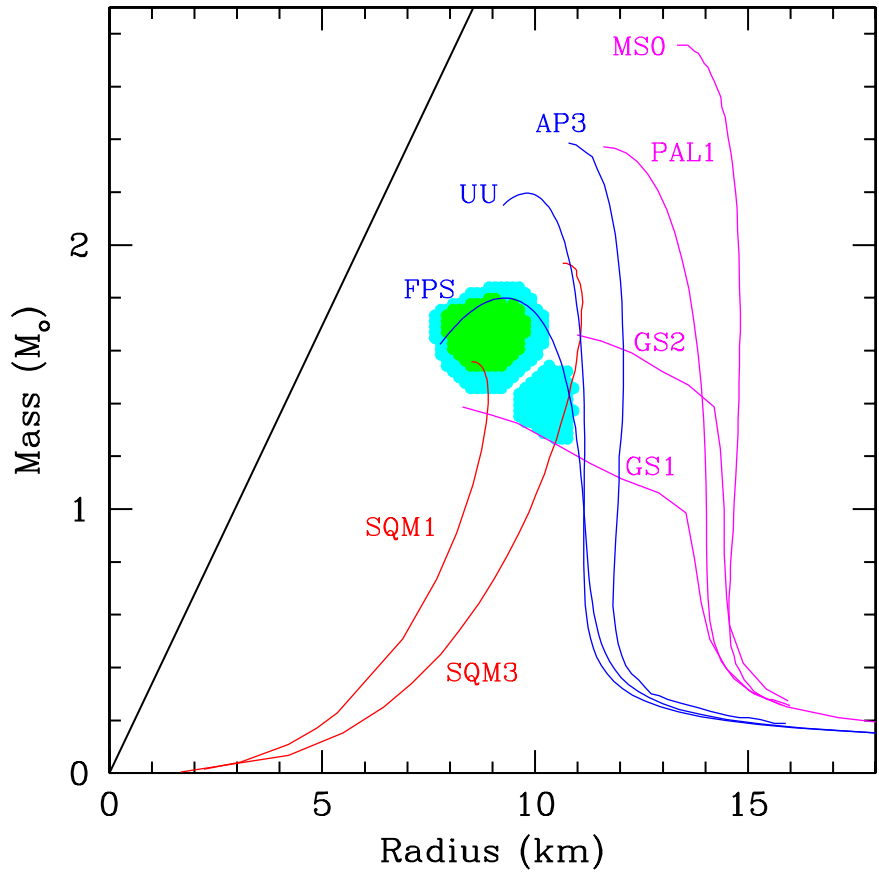} 
\caption{The $1-$ and $2-\sigma$ contours for the mass and radius of the
neutron star in EXO~1745$-$248, for a hydrogen mass fraction of
$X=0$, based on the spectroscopic data during thermonuclear bursts 
combined with a distance measurement to the globular cluster. Neutron 
star radii larger than $\sim 13$~km are inconsistent with the data. 
The descriptions of the various equations of state and the corresponding 
labels can be found in Lattimer \& Prakash (2001). 
}
\label{mr} 
\end{figure}

\end{document}